\pdfoutput=1
\documentclass{article}

\usepackage[final]{nips_2016_whatif}

\usepackage[utf8]{inputenc} 
\usepackage[T1]{fontenc}    
\usepackage{hyperref}       
\usepackage{url}            
\usepackage{booktabs}       
\usepackage{amsfonts}       
\usepackage{nicefrac}       
\usepackage{microtype}      

\usepackage{float}
\usepackage{booktabs}
\usepackage{dcolumn}
\usepackage{multirow}
\usepackage{array}
\newcolumntype{L}[1]{>{\raggedleft\let\newline\\\arraybackslash\hspace{0pt}}m{#1}}
\newcolumntype{C}[1]{>{\centering\let\newline\\\arraybackslash\hspace{0pt}}m{#1}}
\newcolumntype{R}[1]{>{\raggedright\let\newline\\\arraybackslash\hspace{0pt}}m{#1}}

\usepackage{amsmath,amssymb}
\usepackage{amscd,amsthm}

\usepackage{enumitem}
\usepackage{accents}

\usepackage{algorithm}
\usepackage{algorithmic}

\usepackage{graphicx} 

\usepackage{color}
\usepackage[dvipsnames]{xcolor}

\theoremstyle{plain}

\theoremstyle{definition}

\title{
	User Model-Based Intent-Aware Metrics for Multilingual Search Evaluation
}

\author{
	Alexey Drutsa, Andrey Shutovich, Philipp Pushnyakov,  \\
	{\bf Evgeniy Krokhalyov, Gleb Gusev, Pavel Serdyukov} \\
	Yandex;
	16 Leo Tolstoy St., Moscow 119021, Russia \\
	\texttt{\{adrutsa,shutovich,pushnyakov,ekrokhalev,gleb57,pavser\}@yandex-team.ru}
}

\begin{document}

	\maketitle

	\begin{abstract}
Despite the growing importance of multilingual aspect of web search, no appropriate offline metrics to evaluate its quality are proposed so far. At the same time, personal language preferences can be regarded as intents of a query. This approach translates the multilingual search problem into a particular task of search diversification. Furthermore, the standard intent-aware approach could be adopted to build a diversified metric for multilingual search on the basis of a classical IR metric such as ERR.
The intent-aware approach estimates user satisfaction under a user behavior model.
We show however that  the underlying user behavior models is not realistic in the multilingual case, and the produced intent-aware metric do not appropriately estimate the user satisfaction.
We develop a novel approach to build intent-aware user behavior models, which overcome these limitations and convert to quality metrics that better correlate with standard online metrics of user satisfaction.
	\end{abstract}

	\section{Introduction}
	
	\label{sec_Intro}
	
	
	There are many countries whose  population speaks in different languages: a country can have two state languages (e.g., Belgium), have close relations with other countries (e.g.,  Germany), and be subjected to globalization (their citizens actively or passively learn popular international languages)\footnote{\scriptsize http://en.wikipedia.org/wiki/List\_of\_multilingual\_countries\_and\_regions}.
	This forces modern search engines to process queries and documents in different languages for users from the same region, which is known as \emph{multilingual search} or \emph{multilingual aspect} of web search \cite{2005-TALIP-Savoy}.
	
	The language preferences (the need for relevant documents in a particular language) of a user are not always easily deduced from her query to the search engine and may be ambiguous.
	For instance, there are words which  are the same in different languages (like ``table" in English and French), and some named entities have the same meaning in different languages (like  ``cola" and ``CIKM").
	In this paper, we argue that the ambiguity problem of language preferences can be solved by diversification of search results with respect to their languages \cite{2011-AAAI-Chang}.
	Diversification was successfully applied to other types of query intents: navigational/informational \cite{2014-BBIRDB-Sakai}, freshness \cite{2011-CIKM-Styskin}, etc.
	
	A comprehensive overview of various research questions and methodologies  in the field of multilingual search can be found in \cite{2012-SSBM-Peters}, which
	also includes a large survey of CLEF (Conference and Lab of the Evaluation Forum) and its test collections.
	To the best  of our knowledge, there is no study devoted to the evaluation of multilingual search by means of specialized offline metrics. On the face of it, the task of building a multilingual metric may seem straightforward, since a great number of diversified search metrics  exist  \cite{2011-IR-Chapelle,2014-BBIRDB-Sakai}.
	However,  the insufficiency of these metrics becomes apparent, when one applies them to the multilingual search.
	Usually, in the case of the \emph{intent-aware approach}  \cite{2009-WSDM-Agrawal,2011-IR-Chapelle,2014-BBIRDB-Sakai}, quality evaluation is based on relevance assessments of documents assigned for each intent individually.
	However, relevance is essentially independent of the language of a document,
	as the meaning of a document is not supposed to change after its translation into another language.
	Therefore, we can only rely on the relevance labels, which do not account for the language preferences (\emph{universal judgments}).
	For this reason, the state-of-the-art collections like CLEF contain only universal relevance labels~\cite{2012-SSBM-Peters}.
	Therefore, we are faced with the problem of determining the \emph{per-language} relevance \emph{probabilities}  of a document from the document's universal editorial judgment.
	According to the traditional intent-aware approach, we should assume that the relevance probability of a document whose language coincides with the intent (implicit language preference) depends on its universal judgment only, and, if the document language does not coincide with the language preference, the document is  totally irrelevant~\cite{2009-WSDM-Agrawal}.
	This  core principle of the intent-aware approach to diversified retrieval evaluation   could be a pitfall, because,
	in a variety of countries, there is a part of users who can speak or understand two or more languages, though being proficient in these
	languages to a different degree\cite{2012-SSBM-Peters}. The above described approach~\cite{2009-WSDM-Agrawal} can produce a diversified metric, which dos not correctly estimate the satisfaction of such users with search results in different languages.

	In our work, we utilize an intent-aware (IA) approach to make a diversified variant of the offline evaluation metric ERR \cite{2009-CIKM-Chapelle,2011-IR-Chapelle,2014-BBIRDB-Sakai}, which together with its modifications \cite{2011-CIKM-Styskin,2013-SIGIR-Chuklin} are the most popular and well studied offline metrics used in search engine industry and academia \cite{2013-TREC-Collins-Thompson}.
	In order to advance the traditional approach to diversified search evaluation, we modify its underlying user model. Namely, we allow users having one (implicit) language intent to be satisfied by documents in another language. Then, we build a new metric based on this novel intent-aware click model using the technique from \cite{2013-SIGIR-Chuklin}.
	We show experimentally that our extended intent-aware user model outperforms the existing ones in terms of perplexity, and the novel diversified metric (which is based on this IA-model) outperforms the studied offline metrics in terms of their correlation with a set of popular absolute online metrics.

	\section{Framework}
	
	\label{sec_ClickModels}
	We start the development of multilingual metrics from analysis of the click models that underlie the state-of-the-art offline metrics and their diversified variants.
	Metrics investigated in this paper include the state-of-the-art ERR~\cite{2009-CIKM-Chapelle,2013-SIGIR-Chuklin} as a baseline and its different modifications, which are based on the special cases of the Dynamic Bayesian Network (DBN) click model \cite{2009-WWW-Chapelle}.
	
	{\bf Click models.}
	We remind that a click model is a probabilistic model which predicts the user behavior and her clicks on a search engine result page (SERP).
	Particularly, the DBN model assumes \cite{2009-WWW-Chapelle} that a  user examines the document snippets from SERP one by one from top to bottom  and may be attracted by a snippet. If the user is attracted by a snippet, she clicks its URL and, with a certain
	probability, becomes satisfied with the document. If she is not satisfied, she may proceed to the next snippet or stops otherwise. In our work, we restrict ourselves to the simplified version of the DBN model (SDBNs) and add the following constraints
	to align it with the user model underlying ERR metric \cite{2009-WWW-Chapelle}: the user is always attracted by examined snippets, and she never abandons
	search results before having examined all results or having been satisfied with one of them.
	
	In our general framework, we suppose that a user issues a query $q\in\mathcal{Q}$ and examines the first $K$ documents $(d_1,\dots,d_K)$ from the received SERP.
	Let $\mathbb{I}$ be the set of allowed query intents, $I$ be the random variable of the \emph{query's intent} with the value in $\mathbb{I}$, $E_k$ be the random event of \emph{examination} of the $k$-th document $d_k$, and $S_k$ be the random event of \emph{satisfaction} by the $k$-th document $d_k$.
	Then the user behavior is modeled as follows. A user issues a query $q\in\mathcal{Q}$, which has an intent $i\in\mathbb{I}$ with the probability $\mathbf{P}(I\!\!=\!\!i) = p_i$, and starts examining the first document ($E_1$). After the examination of the $k$-th document ($E_k$), she is satisfied ($S_k$) with the probability $\mathrm{pRel}_{i, k}$.
	The described user  behavior is summarized in the following transition probabilities between the states of the variable $I$ and the events $E_k$ and $S_k$ for $k\in\overline{1,K} = \{1,\ldots,K\}$:
	
	\begin{equation*}
	\small
	\begin{split}
	&\mathbf{P}(I\!=\!i \mid \mathrm{init}) = \mathrm{p}_i, \quad
	\mathbf{P}(S_{k} \mid E_{k}, I\!=\!i) = \mathrm{pRel}_{i, k},\: k\in\overline{1,K}, \\
	&\mathbf{P}(E_{k + 1} \mid E_{k}, I\!=\!i) = 1 - \mathrm{pRel}_{i, k}, \: k\in\overline{1,K\!\!-\!\!1},  \quad \mathbf{P}(E_{1})=1
	\end{split}
	\end{equation*}
	for each intent $i\in\mathbb{I}$, where $\mathrm{init}$ is the initial state of the user behavior (before issuing a query)
	and the {\it relevance probability} $\mathrm{pRel}_{i, k}$ defines the probability of satisfaction by the document $d_k$  conditioned by the examination at position $k$.

	According to \cite{2013-SIGIR-Chuklin}, we introduce the following
	additional constraint on the click models in order to build offline evaluation metrics:
	the relevance probability $\mathrm{pRel}_{i,k}$ is determined by the relevance grade of the examined document,  i.e., $\mathrm{pRel}_{i,k} = \mathrm{pRel}_{i}(R_k)$, where, $R_k = R(d_k)\in \{0,\ldots, g_{\max}\}$ is the relevance grade of the $k$-th document $d_k$.
	Hereby, unlike the original SDBN model, it is not an individual parameter for each particular query--document pair.
	The above framework allows us to describe both the SDBN model \cite{2009-WWW-Chapelle,2013-SIGIR-Chuklin}  and its different modifications. In order to obtain a particular click model, one needs to specify the conditional probabilities $\mathrm{pRel}_{i, k}$ and $\mathrm{p}_i$
	for each  $i\in\mathbb{I}$ and $k\in\overline{1,K}$.
	
	{\bf Model-based metrics.}
	Following Chuklin et al. \cite{2013-SIGIR-Chuklin}, we use the state-of-the-art methodology \cite{2009-CIKM-Chapelle,2011-CIKM-Styskin,2013-SIGIR-Chuklin} to build an offline quality metric based on a click model of user behavior.
	The classic effort-based metric on top of the model SDBN is the metric  $\mathtt{ERR}$ \cite{2009-CIKM-Chapelle,2013-SIGIR-Chuklin}.
	To the best of our knowledge, we are the first who proposed to obtain a diversified metric on top of an intent-aware click model.
	The common formula \cite{2009-CIKM-Chapelle} for ERR-family metrics
	is
	
	\begin{equation}\label{eq_ERR_family}
	\small
	ERR= \sum\limits_{i\in\mathbb{I}}\mathrm{p}_i \sum\limits_{k=1}^{K}  \frac{1}{k} \mathrm{pRel}_{i,k}  \prod_{j=1}^{k-1} (1 - \mathrm{pRel}_{i,j})
	\end{equation}
	
	{ \bf Intent awareness.}
	The classical SDBN model \cite{2009-WWW-Chapelle,2013-SIGIR-Chuklin} is intent-agnostic, i.e., there are no query  intents ($|\mathbb{I}|=1$), and its relevance probabilities are independent of the intent $i$: $\mathrm{pRel}_{i,k}(R_k) = p_r(R_k)$ for some map $p_r:\{0,\ldots, g_{\max}\}\rightarrow [0,1]$.
	The simplest approach to introduce an intent awareness into a click model is as follows \cite{2013-ECIR-Chuklin}.  For the intent~$i\in\mathbb{I}$, one introduces the \emph{per-intent} relevance assessments $R^i_k = R^i(d_k)$   to the model's parameters  in place of the (universal) relevance judgments $R_k$  \cite{2011-IR-Chapelle,2014-BBIRDB-Sakai}.
	The editorial judgments \emph{must be obtained}\footnote{In certain cases, there are no per-intent judgments. This is the case of our multilingual study. We will discuss  how we overcome this issue  in Section~\ref{sec_MultLing}.} for each intent individually.
	For instance, for each possible intent of a query and each document,  assessors can be instructed to imagine themselves asking the query with that particular intent, and to ignore the value of the document in the contexts of other possible intents \cite{2011-IR-Chapelle}.
	In such a way, we obtain a modification of the SDBN model, where $\mathrm{pRel}_{i,k} = p_r(R^i_k)$  are substituted as the relevance probabilities (note that the function $p_r$ does not depend on the intent here what will be questioned in the next section).
	The described \emph{intent-aware (IA)} approach is similar to the one generally used to build an offline metric of diversified search from its intent-agnostic variant \cite{2009-WSDM-Agrawal,2011-IR-Chapelle,2014-BBIRDB-Sakai}. Therefore, we will use the same terminology for the described way to introduce intent awareness into a click model.
	

	{\bf  Estimation of model parameters.}
	In order to define the parameters of a model, one either  sets them to default ad-hoc values (i.e., based on intuition only) or fits them from query logs.
	In the first case, for instance, the original ERR metric \cite{2009-CIKM-Chapelle} use the mapping $\mathfrak{R}(g) = \frac{2^g - 1}{2^{g_{\max}} - 1}$, where $g_{\max}$ is the maximum possible relevance grade, thus, e.g., $p_r(g) = \mathfrak{R}(g)$  for intent-agnostic and IA models.
	In the second case, in order to learn a model's parameters (i.e., the relevance probabilities $p_r$ and the intent probability $p_i$), a likelihood function is optimized \cite{2009-WWW-Chapelle,2013-SIGIR-Chuklin}.
	In our work, we do this by means of the BFGS algorithm \cite{opt} (which is a variant of the gradient descend algorithm).

	\begin{figure}
		\includegraphics[width=\textwidth]{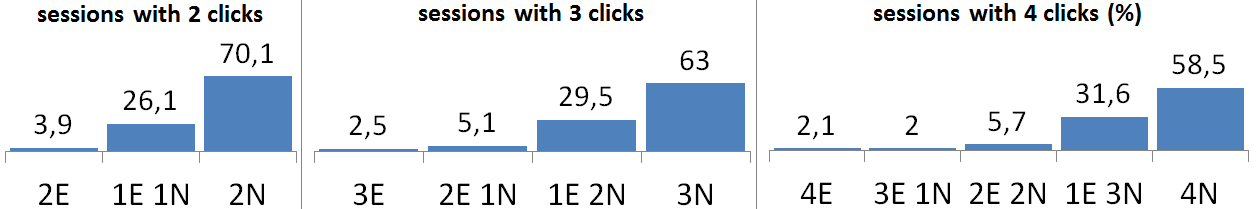}
		\caption{Distribution of sessions with $E$ clicks on English documents and $N$ clicks on documents written in the native language (in \% w.r.t. the total number of sessions with $k$ clicks,  $k\in\overline{2,4}$). }
		\label{fig_ClickDistribution}
	\end{figure}

	\begin{table*}[t]
		\centering
		\caption{The evolution of the relevance probabilities from the source intent-agnostic model via the IA modifications to the EIA modification.}
		\label{tbl_ModelMultiModifications}
		\begin{tabular}{c|c|c|c|c|c|c}
			\multicolumn{1}{c}{  }  & \multicolumn{2}{c}{ \bf  Classic intent-agnostic}
			& \multicolumn{1}{c}{  }  & \multicolumn{2}{c}{ \bf  IA modification}
			& \multicolumn{1}{c}{   }  \\
			\multicolumn{1}{c}{  }  & \multicolumn{2}{c}{  \bf model}
			& \multicolumn{1}{c}{  }  & \multicolumn{2}{c}{  \bf (same params)}
			& \multicolumn{1}{c}{   }  \\
			\cline{2-3}
			\cline{5-6}
			\multicolumn{1}{c}{  $I=$ } &\multicolumn{1}{c}{  $i_1$ } & \multicolumn{1}{c}{  $i_2$ }
			& \multicolumn{1}{c}{   } &\multicolumn{1}{c}{  $i_1$ } & \multicolumn{1}{c}{  $i_2$ }
			& \multicolumn{1}{c}{   } \\
			\cline{2-3}
			\cline{5-6}
			  $L_k=i_1$ &   $p_r(R_k)$ &   $p_r(R_k)$
			&   (1) &   $p^{(1)}_r(R_k)$ &   $p^{(1)}_r(Bad)$
			&   (2) \\
			\cline{2-3}
			\cline{5-6}
			  $L_k=i_2$ &   $p_r(R_k)$  &   $p_r(R_k)$
			&    $\longrightarrow$ &   $p^{(1)}_r(Bad)$  &   $p^{(1)}_r(R_k)$
			&    $\longrightarrow$  \\
			\cline{2-3}
			\cline{5-6}
			
			\multicolumn{1}{c}{  } &
			\multicolumn{1}{c}{  } &
			\multicolumn{1}{c}{  } &
			\multicolumn{1}{c}{  } &
			\multicolumn{1}{c}{  } &
			\multicolumn{1}{c}{  } &
			\multicolumn{1}{c}{  } \\

			\multicolumn{1}{c}{   }  & \multicolumn{2}{c}{  \bf IA modification}
			& \multicolumn{1}{c}{  }& \multicolumn{2}{c}{ \bf  EIA } & \\
			 \multicolumn{1}{c}{   }  & \multicolumn{2}{c}{ \bf  (diff params)}
			& \multicolumn{1}{c}{  }& \multicolumn{2}{c}{ \bf  modification} & \\
			\cline{2-3}
			\cline{5-6}
			 \multicolumn{1}{c}{   } &\multicolumn{1}{c}{  $i_1$ } & \multicolumn{1}{c}{  $i_2$ }
			& \multicolumn{1}{c}{   } &\multicolumn{1}{c}{  $i_1$ } & \multicolumn{1}{c}{  $i_2$ } & \\
			\cline{2-3}
			\cline{5-6}
			   (2) &   $p^{(2)}_{r}(i_1, R_k )$ &   $p^{(2)}_{r}(i_2, Bad )$
			&   (3) &   $p^{(3)}_r(i_1, i_1, R_k )$ &   $p^{(3)}_{r}(i_1, i_2, R_k )$ & \\
			\cline{2-3}
			\cline{5-6}
			   $\longrightarrow$ &   $p^{(2)}_{r}(i_1, Bad )$ &   $p^{(2)}_{r}(i_2, R_k )$
			&     $\longrightarrow$ &   $p^{(3)}_{r}(i_2, i_1, R_k )$ &   $p^{(3)}_r(i_2, i_2, R_k )$ & \\
			\cline{2-3}
			\cline{5-6}
		\end{tabular}
	\end{table*}

	\section{Multilingual intents \& metrics}
	
	\label{sec_MultLing}
	In the case of multilingual search, the space of query intents $\mathbb{I}$ is the set of languages.
	The (universal) editorial judgments\footnote{We consider the state-of-the-art 5-grade scale $\mathcal{R}=\{\hbox{Perfect, Excellent, Good, Fair, Bad}\}$ for the editorial judgments.} $R^i_k$ do not depend on the document's language, since the
	meaning of a document is not supposed to change after its translation into another languages.
	In our work, the baseline is the classical intent-aware approach~\cite{2009-WSDM-Agrawal} (see Sec.~\ref{sec_ClickModels}),
	where the \emph{per-language} relevance judgments $R^i_k, i\in\mathbb{I},$ of a document $d_k$  in
	language $L_k$ is defined as follows.  If the document's language $L_k$ does not coincide with the considered language preference $i$, then the document is naturally treated as totally irrelevant to this intent, i.e., $R^i_k = R_k$, if $i = L_k$, and $R^i_k=0$ ("Bad"), otherwise.
	This approach to introduce intent awareness  
	could be a pitfall
	in the case of the multilingual search by the reasons that we explain further.
	In this paper, \emph{we propose a new intent-aware modification of the SDBN model
		whose relevance  probabilities $\mathrm{pRel}_{i,k}$ depend on both editorial judgments $R_k$ and the combination of the language preference and the language of the document}.

	We modify the SDBN model by increasing, step by step, the degree of freedom of the relevance probabilities $\mathrm{pRel}_{i,k}$. We start from the version presented in the second block of Table~\ref{tbl_ModelMultiModifications}. Here probabilities depend on the editorial judgments solely, if $i = L_k$, and always correspond to the $Bad$ relevance otherwise.
	First, we hypothesize that a user may search for documents in different languages (e.g., her native language and her second language) with different levels of convenience and success.
	\emph{We conclude that the relevance probabilities of a document with the same editorial judgment might be different for different languages}. Therefore, we perform the second step of our modification (denoted by (2) in Table~\ref{tbl_ModelMultiModifications}), where the relevance probabilities are additionally allowed to depend on the intent $i$ in the case $L_k = i$.
	We refer to this version of the IA model as the IA with ``diff params", while we refer to the previous one as the IA  with ``same params".
	
	Second, we remember that there are bilinguals who can speak in or understand two languages. Such a user, while preferring the documents in one language, could be occasionally satisfied by a document written in another language 
	despite that she did not expect that documents in this language would contain any relevant information at the beginning. Such situation could be supported by the observation of user behavior from the query logs of one of the popular search engines  operating in a European country. We plotted the distribution of sessions with $E$ clicks on English documents and $N$ clicks on documents written in the native language in \% w.r.t. the total number of sessions with $k$ clicks, $k\in\overline{2,4}$, in Fig.~\ref{fig_ClickDistribution}. One can see that users click on documents written in both languages in more than $26\%$  of sessions with 2 clicks ($34\%$ and $39\%$   of sessions with $3$ and $4$ clicks respectively). Following this experience,~\emph{we conclude that the relevance probabilities should not always correspond to the $Bad$ relevance in the  case of $i \neq L_k$}. Therefore, we perform the third (final) step of our modification (denoted by (3) in Table~\ref{tbl_ModelMultiModifications}), allowing the relevance probabilities depend both on the language preference $i$ and on the document's language $L_k$ besides the (universal) editorial judgments $R_k$.
	%
	So, in terms of our general framework, we  suppose that 
	$\mathrm{pRel}_{i,k}\!=\!p_r(i,L_k,R_k)$, and obtain a new \emph{Extended Intent-Aware} model (SDBN-EIA).
	

	
	A particular metric of the ERR-family defined by Eq.~\ref{eq_ERR_family} is determined by the parameters $\mathrm{pRel}_{i,k}$ (relevance probabilities)
	and $\mathrm{p}_i$ (intent probabilities), $i\in\mathbb{I},k\in \{1,...,K\}$, that are specified by the click model underlying the metric. 
	For instance, the classical $\mathtt{ERR}$ is based on the model SDBN,  and, thus, it is defined by Eq.~\ref{eq_ERR_family} with default parameters $\mathrm{pRel}_{i,k} = \mathfrak{R}(R_k)$ independent on the query intent (see Sec.~\ref{sec_ClickModels}). Contrariwise, our novel metric with extended intent awareness $\mathtt{ERR\hbox{-}EIA}$ is based on the model SDBN-EIA, and, thus, it  is  defined by Eq.~\ref{eq_ERR_family} with  $\mathrm{pRel}_{i,k} = p_r(i,L_k,R_k)$.
	Note that the modifications that we proposed do not introduce any additional restrictions to the basic click model, but, on the contrary, add more degrees of freedom to it. At the same time, if we were wrong in our assumptions, we would just learn such probabilities from the logs that would transform the extended model to the basic one anyway.
	However, our experiments demonstrate that both the click models and the metrics they underlie them benefit from the additional flexibility. The above modifications improve the metrics, because the better the model predicts user behavior, the better it predicts the user satisfaction, which is determined by Eq.~\ref{eq_ERR_family}.

	\begin{table}
		\centering
		\caption{The average perplexity values for the click models.}
		\label{tbl_ClModelPerplex}
		\begin{tabular}{|l|c|c|l|}
			SDBN modification &   \# of params &  $\mathrm{Perpl}$  & +\% \\
			\hline
			Extended IA (SDBN-EIA)  & 21 & 1.268 & +1.02\% \\
			IA learned ``diff params" & 11 & 1.271 & +0.25\% \\
			IA learned ``same params" & 6 & 1.272 & +3.12\% \\
			Intent-Agnostic learned & 5 & 1.281 & +22.5\% \\
			IA default  & 6 & 1.362 & +17.9\% \\
			Intent-Agnostic default & 5 & 1.441 &  \\
			\hline
		\end{tabular}
	\end{table}

	\section{Experiments}
	
	\label{sec_ClModelCompare}
	
	{\bf Experimental setup.}
	In our experimentation we consider  one of the major web search engines which operates in one of the European countries (15\% of its population have knowledge of foreign languages and 78\% of them speak English).
	In this case,  the space of query intents $\mathbb{I}$ is the set of 2 languages: the \emph{native language} for 99\% of the population of that country and \emph{English language}.
	Since none of the existing collections for multilingual search evaluation \cite{2012-SSBM-Peters} are provided together with any click data (vital for our learning), we have collected click data from the logs of user interactions with the search engine
	during a  six-month
	period in 2013. Then, following Chapelle et al. \cite{2009-CIKM-Chapelle}, we perform the next steps to construct our data set.
	We define a \emph{session} as an event with one query asked by one user, which received a list of results (URLs) and provided a list of clicked
	URLs (unlike in \cite{2009-CIKM-Chapelle}, our session ends with the last action on its SERP).
	We restrict all sessions by the top $5$ URLs of the first result page (i.e., all further clicks were ignored, and, thus, $K = 5$), since, as also explained in \cite{2009-CIKM-Chapelle}, consideration of top 10 positions would lead to a much smaller intersection between query logs and editorial judgments.
	Then, we remove the sessions whose results contain at least one document without an editorial judgment as in \cite{2009-CIKM-Chapelle}.

	Next, specially for multilingual search evaluation, we filter our data as follows. We remove sessions with queries contained non-Latin characters.
	Then, we detect
	\cite{2012-SSBM-Peters}
	the language for each document from the top 5. The sessions with documents written in a language different from the set $\mathbb{I}$ are removed. Finally, we remove sessions whose user's location is outside the country under study. The resulting data set has more
	than  136M  sessions $\mathcal{S}$ and more than 44.8k  unique queries $\mathcal{Q}$.
	Finally, we split the data randomly  into two parts with the ratio $1:9$.
	The smallest part is used as the test data $\mathcal{S}_t$ and
	the largest one serves as the training data $\mathcal{S}_l$. We repeat this procedure $100$ times in order to apply the \emph{paired two-sample t-test} and measure the significance level of the obtained results. Then, each click model, whose parameters need to be learned from clicks, is learned on the training data set $\mathcal{S}_l$ (as described  in Sec.~\ref{sec_ClickModels}).

	{\bf Evaluation of the models.}
	In order to evaluate our models on a test set $\mathcal{S}_t$,  we use a standard \cite{2013-ECIR-Chuklin} averaged perplexity metric $\mathrm{Perpl} = K^{-1}\sum_{k=1}^K p_k$, where, for each position $k\in \overline{1,K}$, we calculate the perplexity $p_k$ from the equality
	
	%
	\begin{equation*}
	\small
	p_k = \prod_{s\in\mathcal{S}_t}\big(\mathbf{P}(C_k \mid s)\big)^{- \frac{C_k(s)}{|\mathcal{S}_t|}} \big(1 - \mathbf{P}(C_k \mid s)\big)^{- \frac{1 - C_k(s)}{|\mathcal{S}_t|}},
	\end{equation*}
	$C_k(s)$ is a binary value that indicates the click event at the $k$-th position in the session $s$, and $\mathbf{P}(C_k \mid s)$ is the  probability of the click on the $k$-th position in the session $s$ under the model.
	The better the model, the lower the value of its perplexity (for an ideal model, it is equal to $1$).
	And also relying on the literature, in order to compute the perplexity gain of a model $M_2$ over a model $M_1$, we use the standard formula $\frac{\mathrm{p}_{M_1} - \mathrm{p}_{M_2}}{\mathrm{p}_{M_1} - 1}\cdot 100\%$.

	In Table~\ref{tbl_ClModelPerplex}, we report the values of the average perplexity $\mathrm{Perpl}$ for all click models under study (see Sec.~\ref{sec_ClickModels} and~\ref{sec_MultLing}). The differences between all pairs of the models are obtained at a high significance level with p-value $< 0.001$.  First, we see that our novel intent-aware model $\mathrm{SDBN\hbox{-}EIA}$
	outperforms  all other studied models (by a margin $>1\%$ over the 2-nd one).
	Second, we see that there is no big difference ($0.25\%$) between the IA models ``same params" and ``diff params" (denoted by (1) and (2) in Table~\ref{tbl_ModelMultiModifications} respectively). Therefore, we conclude that our third  modification (denoted by (3) in Table~\ref{tbl_ModelMultiModifications}) of the relevance probability dependance give more profit than the  second one.
	Third, we see expected results: the order of the models w.r.t.\ $\mathrm{Perpl}$ corresponds to the number of learned parameters, and the models with default parameters have the lowest perplexity by a high margin.
	Finally, we conclude that \emph{our novel intent-aware click model of user behavior outperforms both the state-of-the-art intent-aware click model and intent-agnostic model (i.e., SDBN) in their ability to explain user click behavior}.

	{\bf Evaluation of the metrics.}
	In order to compare the metrics under study (see Sec.~\ref{sec_MultLing}), we calculate correlation between them and some absolute online metrics over configurations.
	We choose this method because it is commonly used \cite{2011-IR-Chapelle,2013-SIGIR-Chuklin}.
	First, we utilize the following absolute online metrics \cite{2009-CIKM-Chapelle,2013-SIGIR-Chuklin}:
	\emph{UCTR} (binary value representing click);
	\emph{MaxRR}, \emph{MinRR}, and \emph{MeanRR} (maximal, minimal, and mean reciprocal ranks of a click in a session); and
	\emph{PLC} (the number of clicks divided by the position of the lowest click).
	Second, \emph{a configuration} is a tuple of a query and the top-$K$ URLs of the ranked documents presented to a user \cite{2009-CIKM-Chapelle,2013-SIGIR-Chuklin}. Our data set has more than 2.1M configurations (i.e., on average, more than $47.7$ configurations per query and more than $63.7$ sessions per configuration).
	We measure the weighted correlation \cite{2009-CIKM-Chapelle} over the configurations in a test data set between
	a model-based offline  and an online metric.
	%
	%

	\begin{table}
		\centering
		\caption{The correlations of the multilingual metrics with online metrics.}
		\label{tbl_MetricCorrelation}
		\begin{tabular}{|l|c|c|c|c|c|}
			\bf ERR modification
			&   UCTR
			&   MaxRR
			&   MinRR
			&   MeanRR
			&   PLC
			\\
			\hline
			Extended IA ({\tt ERR-EIA})&   {\bf 0.318} &   {\bf 0.422} &   {\bf 0.379} &   {\bf 0.386} &   {\bf 0.389} \\
			Intent-Agnostic learned ({\tt ERR}) &   \underline{0.283} &   0.337 &   0.309 &   0.314 &   0.317 \\
			IA learned ``same params" &   0.282 &   \underline{0.409} &   \underline{0.361} &   \underline{0.368} &   \underline{0.372} \\
			Intent-Agnostic default &   0.27 &   0.326 &   0.298 &   0.303 &   0.305 \\
			IA learned ``diff params" &   0.268 &   0.394 &   0.347 &   0.354 &   0.357 \\
			IA default &   0.131 &   0.178 &   0.152 &   0.157 &   0.158 \\
			\hline
			
			%
		\end{tabular}
	\end{table}
	
	We compute the correlations, using the 100-fold sampling of the previous section: we use the same learned parameters of the click models and we calculate the correlation on the test data sets from this sampling. These results are summarized in Table~\ref{tbl_MetricCorrelation} (with p-value $<0.001$).
	First, we see that the ranking of our offline metrics is different with respect to the UCTR and with respect to the other absolute metrics. The difference seems to be caused by the definition of the UCTR, which does not account for clicks unlike the other 4 metrics.
	This finding is in line with the  results from \cite{2009-CIKM-Chapelle,2013-SIGIR-Chuklin}, where the order of the studied metrics is different with respect to the UCTR metric  and with respect to the other online metrics.
	Second, our novel model-based metric $\mathtt{ERR\hbox{-}EIA}$ is the incontestable winner with respect to all absolute metrics.
	Third, we see that the intent-agnostic metric $\mathtt{ERR}$ has the 2-nd place w.r.t.\ UCTR and outperforms some intent-aware metrics w.r.t.\ other absolute metrics.
	Possible explanations of this result, that are discussed in Section~\ref{sec_MultLing}, have encouraged us to study the new intent-aware models. We explain this ``strange" result by peculiarities of multilingual diversification: the presence of bilinguals among the search engine users and a high probability of being satisfied by results in both languages penalize the models,
	where a user, which prefers documents in one language, cannot be satisfied with documents in another one\footnote{The IA models are such models, while intent-agnostic and EIA ones are not (see Sec.~\ref{sec_MultLing}).},

	Finally, we conclude that \emph{our novel model based intent-aware metric outperforms both the state-of-the-art IA metric and the intent-agnostic metric (i.e., ERR) in terms of their correlation with several online metrics}. Moreover, \emph{its use is necessary due to the inferiority of the state-of-the-art IA metric in comparison to the simple intent-agnostic one}.

	\section{Conclusions and future work}
	
	In this paper, we were driven by the need to propose a user model and the corresponding metric which are best suited for the case of multilingual search, motivated by the observation that a considerable portion of users who can understand two
	languages and can be satisfied by documents written in one language, while searching documents in another language. As we demonstrated, the straightforward intent-aware modifications of user models do not take such aspects of this   user behavior into account.
	In passing, first, to the best of our knowledge, we proposed a novel method to obtain new metrics of diversified search, which is based on the conversion of intent-aware
	click models into offline metrics. Second, to the best of our knowledge, we are the first who proposed an offline quality evaluation metric which takes the multilingual aspect of search into account.
	As future work we can,
	first, apply our intent-aware modification of metrics to evaluate diversified search based on other types of query intents, such as freshness, and etc.
	Second,
	we can also experiment with optimization of the click model parameters   by directly maximizing the correlation of the metric it underlies with absolute online metrics.


	
	\small
	
	\bibliographystyle{alpha}
	\bibliography{2016-arxiv-mlcm}
	
\end{document}